Research article                                              **Open Access**

# He$^{2+}$-He charge transfer collisions using a 27-state close-coupled calculation with a diabatic molecular basis

Chanchal Chaudhuri

Address: Department of Physics, University of Calcutta, 92, A. P. C. Road, Kolkata (Calcutta) - 700 009, India

Email: Chanchal Chaudhuri - cchanchal@yahoo.com

## Abstract

A comparative study of two earlier three-state and fifteen-state [Chaudhuri *et al.*, *Pramana-J. Phys.*, **43**, 175 (1994); *ibid.*, *Phys. Rev. A*, **52**, 1137 (1995)] close-coupled treatments of He$^{2+}$-He single and double charge transfer collisions is made in this paper with a larger, 27-state close-coupled calculation. The calculations have been done using the diabatic molecular basis set used in the earlier work extended by adding excited orbitals leading to higher excitation channels up to 3*s* and 3*p*. For such molecular basis functions that go to the correct separated-atom limits used in this work, the present results show that without the inclusion of the electron translation factors (ETFs) the quantitative cross-section calculations up to velocity ~1.1 a.u. (~30 keV/amu) are in good agreement with both experiment and other calculations. This suggests that if ETFs are properly incorporated into the charge transfer collision studies at low energies this diabatic molecular basis can be used for benchmark calculations. With the aid of the Wannier's picture of the ground state correlations, a combined detailed analysis and comparison has been carried out to find a connection with the dynamic two-electron correlation picture in charge transfer collision processes which involve sequential/simultaneous two-electron exchange. If ETFs are included, the model approach of the present work may open up opportunities to investigate dynamic two-electron correlation effect in charge transfer ion-atom collision processes with benchmark accuracy.

## 1. Introduction

We have previously [1,2] made use of a diabatic molecular basis expansion constructed in a similar spirit as in an earlier work [3], to carry out a close-coupled study of single and double charge transfer in $He^{2+}$-$He(1s^2)$ collisions in the impact parameter approximation without electron translation factors (ETFs). The salient features of the results obtained in [1] and [2] can be summarized as follows.

1. For single charge transfer into the ground state, the basis can be truncated very early; there was hardly any difference between the three-state [1] and the fifteen-state [2] results.

2. For resonant double charge transfer, the early truncation was not possible. The results of the fifteen-state calculation agreed very well with experiment [4,5] up to around 10 keV/amu, above which the calculated results dropped rather markedly. It could not be ascertained whether the discrepancy at higher energies was due to the basis set truncation, or due to the neglect of ETFs, or both.

3. The *total* single charge transfer cross-sections obtained by summing over charge transfer channels into the ground state as well as target/projectile excitation channels in the fifteen-state calculations agreed very well with experiment. The individual channel excitations agreed fairly well with a close-coupled calculation done by Fritsch [6] using travelling atomic orbitals, and showed a very similar energy variation.

In our earlier work [2] it was noted that in the three-state calculation very little configuration interaction (CI) had been included in the wave function (of the collision complex), whereas a considerable amount of CI went in the fifteen-state basis. Hence, the *dynamic electron correlation* phenomena could be involved in two-electron rearrangement collisions. This phenomena failed to appear in the 3-state calculations, but it should show up in the fifteen-state calculations. However, discussion of this point in [2] was limited to the degree of success (or lack of it) in predicting *total charge transfer cross-sections* alone. This article presents a theoretical investigation of the $He^{2+}$-$He(1s^2)$ charge transfer collision system with a 27-state diabatic molecular basis-set including excitation channels up to ($3s$, $3p$) without ETFs, (the consistency of the exclusion of ETFs in the 3-state [1], 15-state [2] and 27-state calculations throughout is preserved). From a detailed comparative study among the 3-state [1], 15-state [2] and 27-state, the present work focuses on the investigation of the *dynamic electron correlation* in the two-electron ion-atom collision system $He^{2+}$-$He(1s^2)$ with the help of the Wannier's picture of the ground state correlations. For this purpose, the present results are compared mainly with the radial and angular electron correlation in a two-electron atomic system in Wannier's [7] study.

This work is carried out mainly with an aim that, whether or not, present approach based on the close-coupled 27-state diabatic molecular basis calculations without ETFs can be applied to investigate complicated and interesting *dynamic electron correlation* behaviour in a two-electron ion-atom collision system. In other view, if one can choose to study the phenomena of dynamic correlated behaviour of electronic motions in ion-atom/molecule collisions with computational ease (i.e., neglecting ETFs) using the model approach of the present work, or not.

## 2. Theoretical method

The total electronic Hamiltonian of $He_A + He_B^{2+}$ system is given by (atomic units are used throughout)

$$H_{el} = H_0 + W + 4/R$$

$$\text{where } H_0 = \left(-\frac{\nabla_1^2}{2} - \frac{2}{|\vec{r}_1 - \vec{R}_A|} - \frac{2}{|\vec{r}_1 - \vec{R}_B|}\right) + \left(-\frac{\nabla_2^2}{2} - \frac{2}{|\vec{r}_2 - \vec{R}_A|} - \frac{2}{|\vec{r}_2 - \vec{R}_B|}\right); W = \frac{1}{|\vec{r}_1 - \vec{r}_2|}$$

(1)

Here $\vec{r}_1$ and $\vec{r}_2$ are the position vectors of electron 1 and electron 2 relative to the origin (center of mass of the system), respectively. $R = |\vec{R}| = |\vec{R}_A - \vec{R}_B|$ where $\vec{R}_A$ and $\vec{R}_B$ are the position vectors of the $He_A$ and $He_B$ nuclei from relative to the origin, respectively. $W$ is the electron-electron interaction term and $4/R$ is the internuclear potential energy.

First we expand the total wavefunction of the collision system in a body-fixed truncated molecular basis,

$$\Psi(\vec{r}, \vec{R}(t)) = \sum_k c_k(t) \psi_k(\vec{r}, R)$$

(2)

Then in the semiclassical impact-parameter approximation, the time dependent Schrödinger equation takes the form (for details see [3])

$$i\frac{dc_j}{dt} = \sum_k (H_{jk} + P_{jk} + Q_{jk})c_k$$

(3)

With the same sprit as in [3], for use in eqns.(2) and (3) we define the diabatic molecular basis as follows

$$\psi(\vec{r}, R) = \underline{U}(R = \infty) \underline{S}(R) \phi(r, \zeta(R = \infty); R)$$

(4)

where $\psi$ and $\phi$ are column vectors, $\zeta$ is the orbital exponent, $\underline{S}R$ is the Schmidt orthonormalization matrix and $\underline{U}$ diagonalizes $\langle \underline{S}\phi|H_{el}|\underline{S}\phi\rangle$ at large R. Table 1 gives the basis functions $\phi_k$'s used to construct the diabatic molecular wavefunctions $\psi_k$'s (cf. eqn.(3) in [2]), whereas Table 2 (data in the fourth column are collected from Ref.[8]) gives the separated-atom limits of the former. As before, the molecular orbitals were built up with a minimal Slater basis.

These have been adequately described earlier [1,2]; suffice it to say that in this paper the diabatic basis has been extended to include up to (3s,3p) excitations in the separated-atom limit which form 27-state basis-set expansions. For computational ease and to keep the consistency in

**Table 1: Basis functions (normalized) used to construct the diabatic molecular wavefunctions. The {*A*, *B*} denotes the spatially symmetric (singlet) combination $(A_1B_2 + A_2B_1)(\alpha_1\beta_2 - \alpha_2\beta_1)$ where $\alpha$, $\beta$ are "up" and "down" spin functions and *A*, *B* are spatial MO's.**

Singlet g-state

$$\phi_1 = \tfrac{1}{\sqrt{2}}[\sigma_g(1s)^2 - \sigma_u(1s)^2]$$

$$\phi_2 = \tfrac{1}{\sqrt{2}}[\sigma_g(1s)^2 + \sigma_u(1s)^2]$$

$$\phi_3 = \tfrac{1}{2}[\{\sigma_g(1s)\pi_g(2p_+)\} - \{\sigma_u(1s)\pi_u(2p_+)\}]$$

$$\phi_4 = \tfrac{1}{2}[\{\sigma_g(1s)\pi_g(2p_+)\} + \{\sigma_u(1s)\pi_u(2p_+)\}]$$

$$\phi_5 = \tfrac{1}{2}[\{\sigma_g(1s)\sigma_g(2s)\} - \{\sigma_u(1s)\sigma_u(2s)\}]$$

$$\phi_6 = \tfrac{1}{2}[\{\sigma_g(1s)\sigma_g(2s)\} + \{\sigma_u(1s)\sigma_u(2s)\}]$$

$$\phi_7 = \tfrac{1}{2}[\{\sigma_g(1s)\sigma_g(2p_0)\} - \{\sigma_u(1s)\sigma_u(2p_0)\}]$$

$$\phi_8 = \tfrac{1}{2}[\{\sigma_g(1s)\sigma_g(2p_0)\} + \{\sigma_u(1s)\sigma_u(2p_0)\}]$$

$$\phi_9 = \tfrac{1}{2}[\{\sigma_g(1s)\sigma_g(3s)\} - \{\sigma_u(1s)\sigma_u(3s)\}]$$

$$\phi_{10} = \tfrac{1}{2}[\{\sigma_g(1s)\sigma_g(3s)\} + \{\sigma_u(1s)\sigma_u(3s)\}]$$

$$\phi_{11} = \tfrac{1}{2}[\{\sigma_g(1s)\sigma_g(3p_0)\} - \{\sigma_u(1s)\sigma_u(3p_0)\}]$$

$$\phi_{12} = \tfrac{1}{2}[\{\sigma_g(1s)\sigma_g(3p_0)\} + \{\sigma_u(1s)\sigma_u(3p_0)\}]$$

$$\phi_{13} = \tfrac{1}{2}[\{\sigma_g(1s)\pi_g(3p_+)\} - \{\sigma_u(1s)\pi_u(3p_+)\}]$$

$$\phi_{14} = \tfrac{1}{2}[\{\sigma_g(1s)\pi_g(3p_+)\} + \{\sigma_u(1s)\pi_u(3p_+)\}]$$

Singlet u-state

$$\phi_{15} = \tfrac{1}{\sqrt{2}}\{\sigma_g(1s)\sigma_u(1s)\}$$

$$\phi_{16} = \tfrac{1}{2}[\{\sigma_g(1s)\pi_u(2p_+)\} - \{\sigma_u(1s)\pi_g(2p_+)\}]$$

$$\phi_{17} = \tfrac{1}{2}[\{\sigma_g(1s)\pi_u(2p_+)\} + \{\sigma_u(1s)\pi_g(2p_+)\}]$$

$$\phi_{18} = \tfrac{1}{2}[\{\sigma_g(1s)\sigma_u(2s)\} - \{\sigma_u(1s)\sigma_g(2s)\}]$$

$$\phi_{19} = \tfrac{1}{2}[\{\sigma_g(1s)\sigma_u(2s)\} + \{\sigma_u(1s)\sigma_g(2s)\}]$$

$$\phi_{20} = \tfrac{1}{2}[\{\sigma_g(1s)\sigma_u(2p_0)\} - \{\sigma_u(1s)\sigma_g(2p_0)\}]$$

$$\phi_{21} = \tfrac{1}{2}[\{\sigma_g(1s)\sigma_u(2p_0)\} + \{\sigma_u(1s)\sigma_g(2p_0)\}]$$

$$\phi_{22} = \tfrac{1}{2}[\{\sigma_g(1s)\sigma_u(3s)\} - \{\sigma_u(1s)\sigma_g(3s)\}]$$

$$\phi_{23} = \tfrac{1}{2}[\{\sigma_g(1s)\sigma_u(3s)\} + \{\sigma_u(1s)\sigma_g(3s)\}]$$

$$\phi_{24} = \tfrac{1}{2}[\{\sigma_g(1s)\sigma_u(3p_0)\} - \{\sigma_u(1s)\sigma_g(3p_0)\}]$$

$$\phi_{25} = \tfrac{1}{2}[\{\sigma_g(1s)\sigma_u(3p_0)\} + \{\sigma_u(1s)\sigma_g(3p_0)\}]$$

$$\phi_{26} = \tfrac{1}{2}[\{\sigma_g(1s)\pi_u(3p_+)\} - \{\sigma_u(1s)\pi_g(3p_+)\}]$$

$$\phi_{27} = \tfrac{1}{2}[\{\sigma_g(1s)\pi_u(3p_+)\} + \{\sigma_u(1s)\pi_g(3p_+)\}]$$

the 3-state, 15-state and 27-state calculations throughout, the ETFs have been neglected in the present calculation. The center of mass coincides with the center of charge for a symmetric system, and it has been chosen as the origin of the co-ordinates in [1,2] and in the present work. In this work the equations to be solved are (using the notations and procedure of [2])

$$i\frac{dC_j}{dt} = \sum_{k \neq j}^{1,27} C_k (H_{jk} + Q_{jk}) \exp[i\int (H_{jj} - H_{kk})dt] \tag{5}$$

where the Hamiltonian matrix elements $H_{jk} = <\psi_j|H_{el}|\psi_k>$ and the rotational coupling matrix elements $Q_{jk} = \frac{vb}{R^2} <\psi_j | -i\frac{\partial}{\partial \theta} | \psi_k >$.

Because of the choice of the origin at the centre of the internuclear line, the *g* and *u* subsets of the coupled eqn.(5) separate, and they were solved by the Bulirsch-Stoer method [9], unitarity being preserved to within 4–5 parts in $10^4$ (or better). The different charge transfer channel probabilities were then determined using similar formulas as (8) in [2], whereupon the different channel cross-sections were obtained by standard methods.

**Table 2: Separated-atom behavior of the basis states.**

| Basis functions | Separated-atom limit | Energy (a.u.) calculated at R = 50 a.u. | Energy (a.u.) from Ref. [8] |
|---|---|---|---|
| $\phi_1$ | $\{1s_A(1)1s_B(2)\}$ | -3.98000 | -3.97963 |
| $\phi_2$ | $[1s_A(1)1s_A(2) + 1s_B(1)1s_B(2)]$ | -2.85176 | -2.90335 |
| $\phi_3$ | $\{1s_B(1)2p_{+A}(2)\} - \{1s_A(1)2p_{+B}(2)\}$ | -2.47999 | -2.47978 |
| $\phi_4$ | $\{1s_A(1)2p_{+A}(2)\} - \{1s_B(1)2p_{+B}(2)\}$ | -2.12229 | -2.12363 |
| $\phi_5$ | $\{1s_A(1)2s_B(2)\} + \{1s_B(1)2s_A(2)\}$ | -2.47442 | -2.47978 |
| $\phi_6$ | $\{1s_A(1)2s_A(2)\} + \{1s_B(1)2s_B(2)\}$ | -2.14028 | -2.14576 |
| $\phi_7$ | $\{1s_B(1)2p_{0A}(2)\} - \{1s_A(1)2p_{0B}(2)\}$ | -2.48001 | -2.47978 |
| $\phi_8$ | $\{1s_A(1)2p_{0A}(2)\} - \{1s_B(1)2p_{0B}(2)\}$ | -2.12226 | -2.12363 |
| $\phi_9$ | $\{1s_B(1)3s_A(2)\} + \{1s_A(1)3s_B(2)\}$ | -2.19596 | -2.20201 |
| $\phi_{10}$ | $\{1s_A(1)3s_A(2)\} + \{1s_B(1)3s_B(2)\}$ | -2.06322 | -2.06109 |
| $\phi_{11}$ | $\{1s_B(1)3p_{0A}(2)\} - \{1s_A(1)3p_{0B}(2)\}$ | -2.20189 | -2.20201 |
| $\phi_{12}$ | $\{1s_A(1)3p_{0A}(2)\} - \{1s_B(1)3p_{0B}(2)\}$ | -2.05084 | -2.05495 |
| $\phi_{13}$ | $\{1s_B(1)3p_{+A}(2)\} - \{1s_A(1)3p_{+B}(2)\}$ | -2.20140 | -2.20201 |
| $\phi_{14}$ | $\{1s_A(1)3p_{+A}(2)\} - \{1s_B(1)3p_{+B}(2)\}$ | -2.05377 | -2.05495 |
| $\phi_{15}$ | $\{1s_A(1)1s_{+A}(2)\} - \{1s_B(1)1s_B(2)\}$ | -2.85161 | -2.90335 |
| $\phi_{16}$ | $\{1s_A(1)2p_{+B}(2)\} + \{1s_B(1)2p_{+A}(2)\}$ | -2.47999 | -2.47978 |
| $\phi_{17}$ | $\{1s_A(1)2p_{+A}(2)\} + \{1s_B(1)2p_{+B}(2)\}$ | -2.12229 | -2.12363 |
| $\phi_{18}$ | $\{1s_B(1)2s_A(2)\} - \{1s_A(1)2s_B(2)\}$ | -2.47452 | -2.47978 |
| $\phi_{19}$ | $\{1s_A(1)2s_A(2)\} - \{1s_B(1)2p_B(2)\}$ | -2.14051 | -2.14576 |
| $\phi_{20}$ | $\{1s_A(1)2p_{0B}(2)\} + \{1s_B(1)2p_{0A}(2)\}$ | -2.47989 | -2.47978 |
| $\phi_{21}$ | $\{1s_A(1)2p_{0A}(2)\} + \{1s_B(1)2p_{0B}(2)\}$ | -2.12214 | -2.12363 |
| $\phi_{22}$ | $\{1s_B(1)3s_A(2)\} - \{1s_A(1)3s_B(2)\}$ | -2.19609 | -2.20201 |
| $\phi_{23}$ | $\{1s_A(1)3s_A(2)\} - \{1s_B(1)3s_B(2)\}$ | -2.06317 | -2.06109 |
| $\phi_{24}$ | $\{1s_B(1)3p_{0A}(2)\} + \{1s_A(1)3p_{0B}(2)\}$ | -2.20195 | -2.20201 |
| $\phi_{25}$ | $\{1s_A(1)3p_{0A}(2)\} + \{1s_B(1)3p_{0B}(2)\}$ | -2.05078 | -2.05495 |
| $\phi_{26}$ | $\{1s_B(1)3p_{+A}(2)\} + \{1s_A(1)3p_{+B}(2)\}$ | -2.20140 | -2.20201 |
| $\phi_{27}$ | $\{1s_A(1)3p_{+A}(2)\} + \{1s_B(1)3p_{+B}(2)\}$ | -2.05382 | -2.05495 |

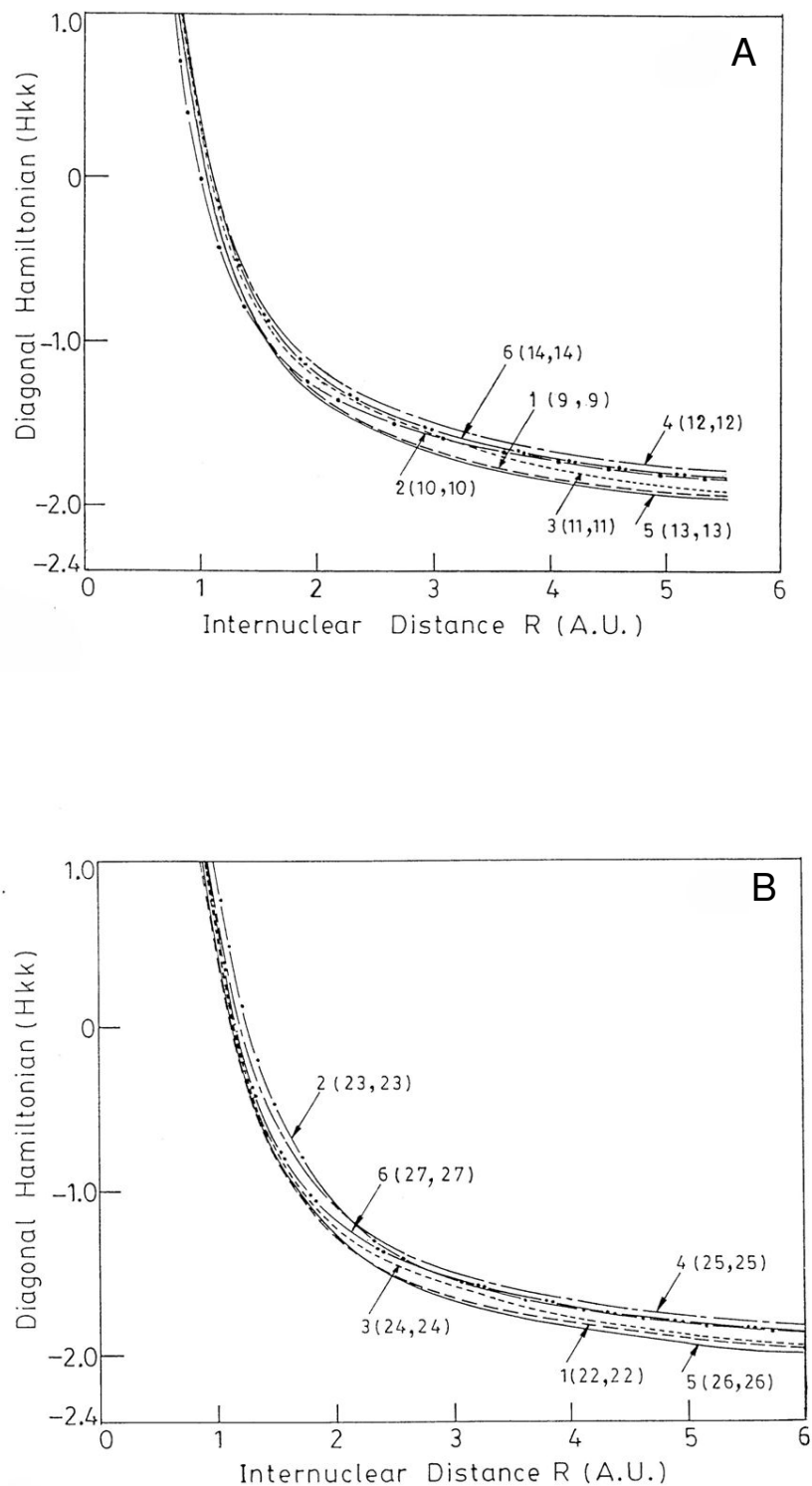

**Figure 1**
**(A) Some diabatic diagonal Hamiltonian matrix elements ($H_{kk}$) for new $\Sigma_g$-states plotted against internuclear distance R**. (B) Same for new $\Sigma_u$-states plotted against internuclear distance R. For numbering of the states |k>, see Table 1.

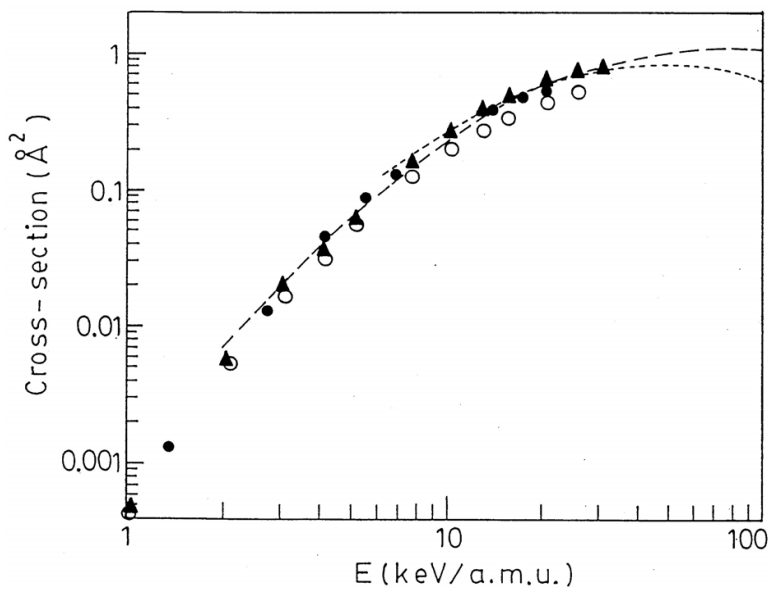

**Figure 2**
**Single charge transfer cross sections for reaction (6a) plotted as function of laboratory energy**. Dark triangles: present calculation (27-state basis); open circles: 15-state basis calculation [2]; dashed line: 3-state calculation [1]; dotted line: Fulton and Mittleman [10]; dark circles: Afrosimov's data [4].

## 3. Results

The following charge transfer reactions have been studied:

(i) Single capture:

$$He_A(1s^2) + He_B^{2+} \to He_A^+(1s) + He_B^+(1s) \tag{6a}$$

$$He_A^+(nl) + He_B^+(1s) \tag{6b}$$

$$He_A^+(1s) + He_B^+(nl) \tag{6c}$$

[(*nl*)'s are excited-state configurations]

(ii) Resonant double capture:

$$He_A(1s^2) + He_B^{2+} \to He_A^{2+} + He_B(1s^2) \tag{7}$$

at laboratory energies up to ~30 keV/amu. As in [2], *simultaneous* target and projectile excitation channels have been ignored.

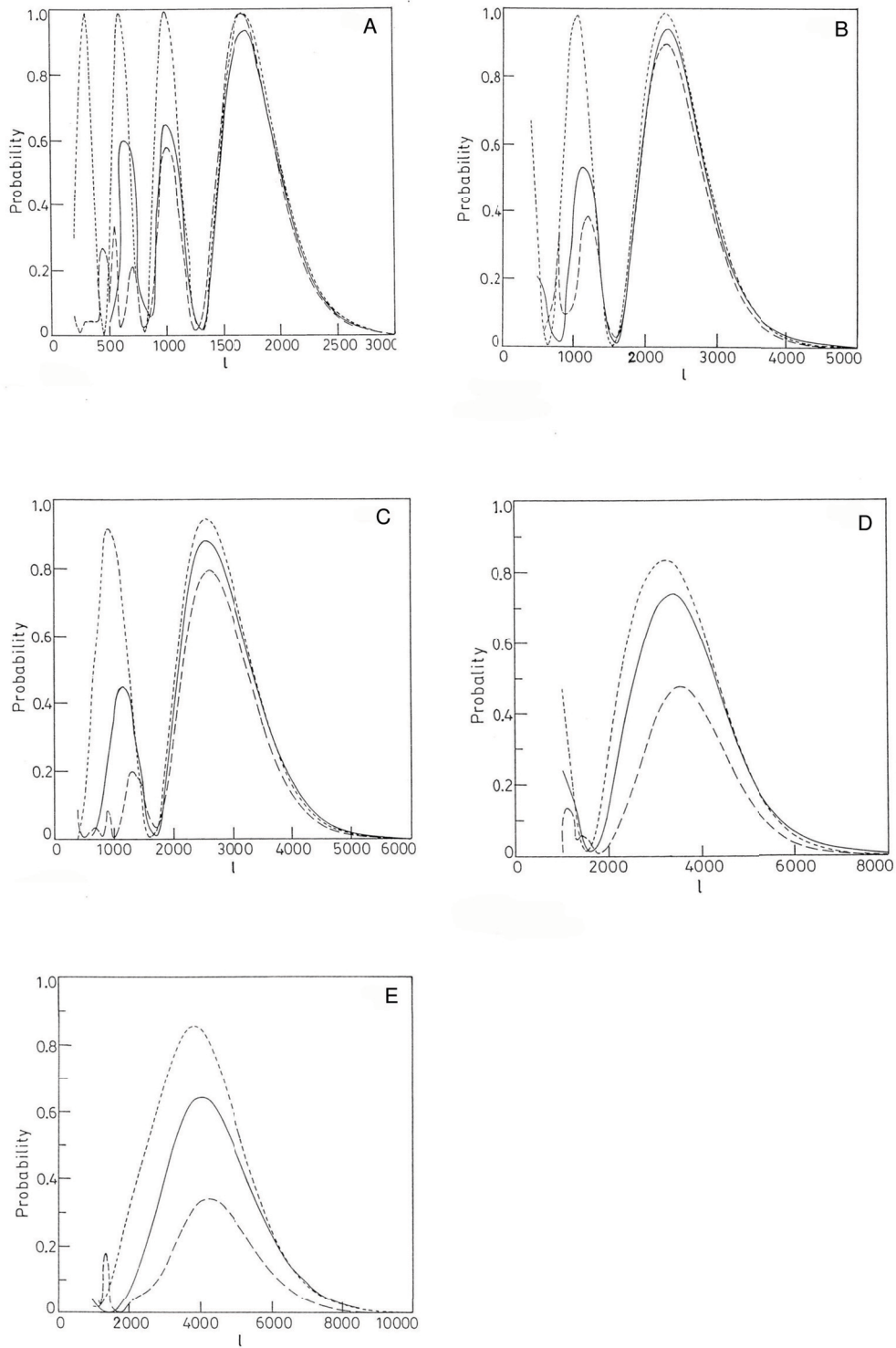

**Figure 3**
**(A-E). Comparison of resonant double charge transfer probabilities among 27-state calculation (solid line), 15-state calculation (dashed line) and 3-state calculation (dotted line) at energies 2, 5, 7.5, 15 and 25 keV/amu, respectively**.

In [2] we showed the R-variation of some diabatic Hamiltonian matrix elements $H_{jk}$ involving states covering up to 2s-2p excitations. The diabatic Hamiltonians involving the new states (covering up to 3s-3p excitations) considered in this paper have very similar R-variations, as shown in Figs. 1. A qualitative and quantitative comparison of the diagonal Hamiltonian matrix elements ($H_{kk}$) of the six new *g* state (Fig. 1A) and six new *u* states (Fig. 1B of the present calculation with those of the 3-state [1] and 15-state calculation [2] can be done from scale of the abscissa and ordinate. The plots of $H_{kk}$ in Figs.1 supplement to those of our earlier work [1,2]. The rotational coupling elements $Q_{jk}$ are remarkably free from spurious asymptotic behaviour, as already shown earlier [3].

Fig.2 shows the ground-state single capture (channel (6a) above) cross-sections, as obtained in the 3-state, 15-state and the 27-state calculations. Also shown for comparison are the experimental data of Afrosimov *et al*. [4,5]. It is clear that for this reaction channel, early truncation of the diabatic basis after three states gives sufficiently accurate cross-sections, as had already been noted in [1,2]. Inclusion of excited states, up to (2s, 2p) in [2] and up to (3s, 3p) in the present work seems to produce little effect gradually, though a certain *oscillatory* nature appears in the respective contributions. This point will be discussed again shortly. Cross-sections for the *total single capture*, summed over reaction channels (6a – 6c), differ very little from [2] and are not shown.

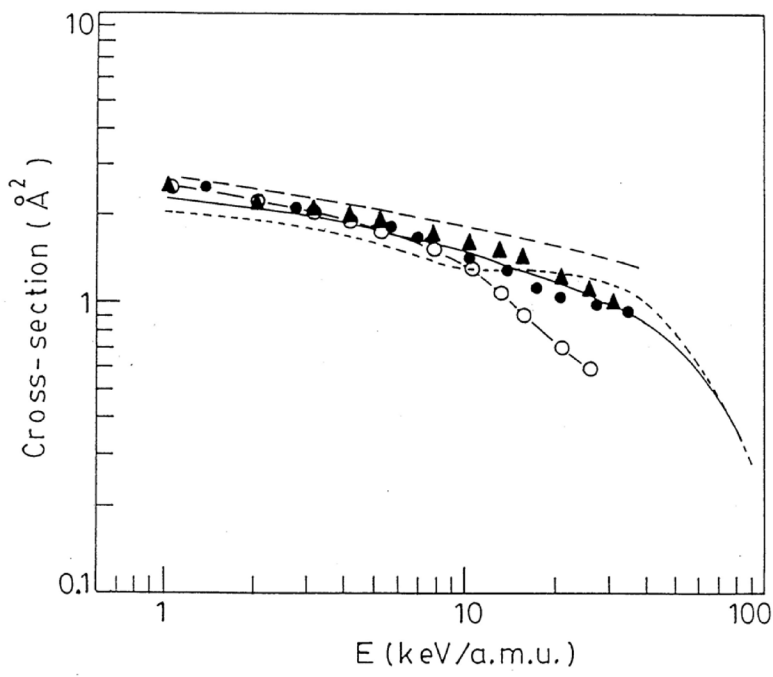

**Figure 4**
**Resonant double charge transfer cross sections for reaction channel (7) plotted against energy**. Dark triangles: present calculation (27-state basis); open circles: 15-state basis calculation [2]; dashed line: 3-state calculation [1]; solid line: Kimura [11]; dotted line: Fulton and Mittleman [10]; dark circles: Afrosimov's data [5].

Figs. 3A–E display the probabilities ($P_k$, k = 3, 15 and 27) against orbital angular momentum *l* for the resonant double capture channel for 3-state, 15-state and 27-state calculations at a few different energies (2, 5, 7.5, 15 and 25 keV/amu). The impact parameter *b* has been equated to the orbital angular momentum *l* via the semiclassical equivalence relationship $b = (l + \frac{1}{2})\hbar/\mu v$, where $\mu$ is the reduced mass of the system and *v* the relative velocity. The dependence of the *P*-values upon the size of the diabatic basis chosen is discussed in the next section in the light of Wannier's theory of electron correlation in a two-electron system.

Fig.4 represents the resonant double charge transfer cross-sections obtained in the 27-state calculations; also shown for comparison are the 3-state and 15-state results already exhibited in [2], the data of Afrosimov *et al.* [5], the three-state travelling atomic orbital basis calculations of Fulton and Mittleman [10] and the travelling adiabatic molecular basis calculations of Kimura [11]. The next section illustrates that if dynamic electron correlation is properly taken into account by configuration interaction then charge transfer reactions in ion-atom collisions can be very well described in close-coupled calculations using these molecular basis functions used in the present work without inclusion of momentum translation factors (ETFs), unless one requires benchmark results.

## 4. Implication of dynamic electron correlation

From Fig.4 it is evident that the 3-state, the 15-state and the 27-state calculations agree progressively better with the experiment on resonant double charge transfer to the ground state. For a one-electron system, the only error involved lies in truncating the sum – dubbed the "resolution of the identity" —

$$1 = \sum_{n=1}^{\infty} |n\rangle\langle n| \qquad (8)$$

after a *finite* number of terms that introduces the error, commonly known as the 'size effect', in close-coupled calculations.

At this point note that with any arbitrary choice of basis we can always resort to a "brute force" technique to eliminate, or at least minimize, this error, but to keep within a reasonable size, we must see that the chosen basis does exhibit some properties of the wave function of the system under study. For a two-electron system, an additional property that comes into the picture is *electron correlation*. It was in this context that the usefulness of a configuration interaction to incorporate a certain degree of electron correlation was discussed in our earlier work [2] where we showed that, in our diabatic basis, early truncation of the series involves a sacrifice of configuration interaction and can lead to neglect of electron correlation. (This was, of course, a demonstration of a point already made by Smith [12].)

It is notable, however, that when quantum chemistry deals with electron correlation and incorporating the same in a one-electron orbital picture via configuration interaction, it actually means *static correlation*, defining the "correlation energy" as the difference between the exact (non-relativistic) energy and the Hartree-Fock energy, the latter being the best possible single-configuration energy within the orbital picture. In contrast, in a collision process we are actually interested in the *dynamic correlation*, i.e. correlated behaviour of two electrons involved in a collision process. For the present purposes it is defined as to be the "behavioural pattern" of the two active electrons in an atomic collision under circumstances when they seem to move "concertedly"; obviously one can generally expect such patterns in all *two-electron processes*.

Let us recall the Coulomb potential for the two electrons in Helium. Expressed in the hyperspherical coordinates, it reads (see e.g. [13])

$$V(R,\alpha,\theta_{12}) = \frac{e^2}{R} C(\alpha,\theta_{12}),$$

$$C(\alpha,\theta_{12}) = -\frac{2}{\cos\alpha} - \frac{2}{\sin\alpha} + \frac{1}{\sqrt{1-\cos\theta_{12}\sin 2\alpha}}$$

Here 'hyperradius' $R = \sqrt{r_1^2 + r_2^2}$, 'hyperangle' $\alpha = \tan^{-1}(r_2/r_1)$ and $\theta_{12}$ is the angle between the position vectors $\vec{r}_1$ and $\vec{r}_2$ for the two electrons. The coulomb potential $V(R, \alpha, \theta_{12})$ contains the radial correlation through $R$ and $\alpha$, and the angular correlation through $C(\alpha, \theta_{12})$. Fig. 5 illustrates the potential surface function $C(\alpha, \theta_{12})$ on the $(\alpha, \theta_{12})$-plane; the shaded area shows the "Wannier ridge". Wannier [7] showed that a pair of electrons, which travel along trajectories diverging about this ridge, i.e. move correlatedly such that $r_1 \approx r_2$ and $\hat{r}_1 \approx -\hat{r}_2$ i.e., keeping the opposite sides of the nucleus, would emerge *together* from the atom at near-threshold energies.

As a consequence of the Wannier picture, therefore, it is quite expected that when a bare He nucleus approaches the He atom, correlation effects should show up in the double charge transfer dynamics, *provided the following conditions are satisfied*:

(4.1) the two electrons attached to the target have been moving preferentially along the Wannier ridge,

(4.2) the impact velocity is of the order of (more exactly, *neither much smaller nor much larger than*) the orbital velocity of the electrons, and

(4.3) the impact parameter lies within a few (typically ~3) times the hyper-radius R.

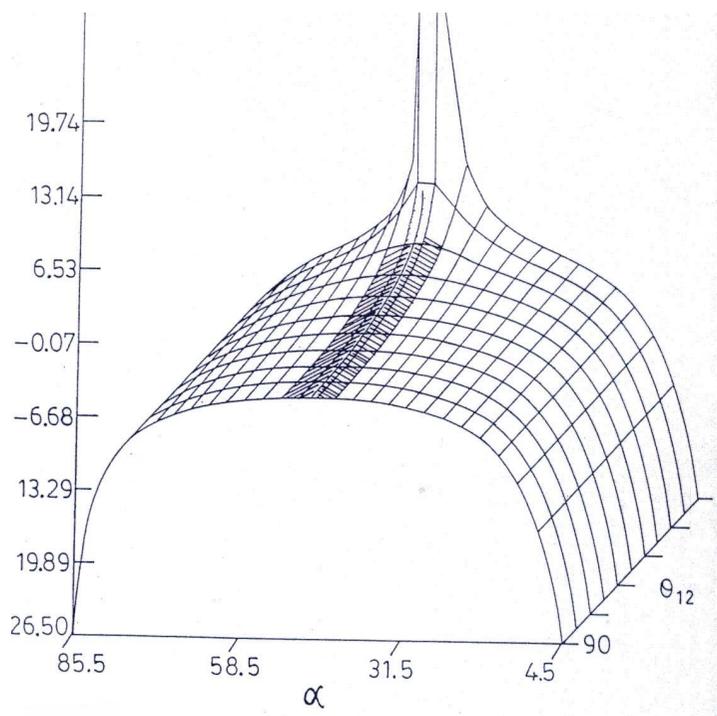

**Figure 5**
**Plot displays the function C($\alpha$, $\theta_{12}$) (vertical axis) versus $\alpha$ and $\theta_{12}$; the shaded area shows the "Wannier ridge"**.

In the ground-state Helium atom the two electrons are strongly correlated, and the condition (4.1) is well satisfied. As a consequence of (4.1), static correlation would be carried over as dynamic correlation. Regarding (4.2), note that two-electron capture in an ion-atom collision can occur (conceptually) either as a *sequential*, two-step process or as a *simultaneous* double capture process [14], and condition (4.2) is essentially the obvious requirement that electron correlation would be important only if sequential capture does not predominate over simultaneous capture of the two electrons. Apart from this, condition (4.2) also follows from the fact that at energies much above threshold, the Wannier picture of radial correlation ceases to hold. Condition (4.3) states that the projectile ion must be moving close enough to the target atom so that, during capture, the two electrons can pass over without "riding" too high up the ridge; in the latter eventuality, the probability is high that the two-electron system would 'slide down' one of the valleys ($\alpha$ = 0 or $2\pi$) i.e., the target would lose only one electron. From Fig.5 we see that a rough estimate for this can be made by requiring angular correlation that after capture, $\theta_{12} \geq 18°$, i.e., the impact parameter lies within ~3 times the hyper-radius of the target atom. It is obvious that fulfillment (or not) of this condition can depend crucially on the choice of the orbital basis; for example, inclusion of excited orbitals offers the collision system to expand to a larger size. Whether or not, it really does expand is a matter of the actual dynamic calculation of the collision system. If any one of these conditions is violated, there would be little or no effect of electron correlation on the double charge transfer channel.

To incorporate such investigations in a theoretical study, one has two alternatives:

(1) Use basis functions of a type that involves the interelectronic distance *explicitly*, or

(2) Use a conventional (atomic or molecular) orbital basis. It is obvious that a much larger basis set would be required here than in (1) above to achieve the same degree of configuration interaction.

Basis functions of type (1) include Hylleraas-type functions, Jacobi polynomials (using hyperspherical coordinates), etc. Although the methodology of (1) above is a more "direct" approach to study electron correlation, its use in atomic collision problems is beset with mathematical difficulties, and (to our knowledge) it has not been attempted yet for the problem of electron transfer in ion-atom collisions. This article attempts that through the inclusion of CI in a close-coupled treatment using the diabatic molecular basis used in this work, one should be able to establish contact with the Wannier picture outlined above and may glean an idea of the dynamic aspect of electron correlation in a double charge transfer ion-atom collision.

As already noted, the differences between the smaller-basis and the larger-basis calculations in this two-electron system stem from *two factors*; the basis size effect and the electron correlation effect. The problem of separating out the signatures of the 'size effect' from those of the 'dynamic correlation effect' in the calculated results has no unique solution, but the following procedure which could be taken as an indication, has been adopted for the present study. Eqn.(8) is, of course, equivalent to the equation (1) of [2] and the 'basis expansion equation' (2) of the present calculation. In the present calculation, eqn.(8) is translated as *the unitarity condition*

$$\sum_{k=1}^{27}|c_k(t)|^2 = 1$$

where, $c_k(t) = \langle \psi_k | \Psi \rangle$ ($\psi_k \equiv |n\rangle$).

Now let us define the quantities

$$S_3 = \sum_{k}^{3}|c_k(t \to \infty)|^2, \quad S_{15} = \sum_{k}^{15}|c_k(t \to \infty)|^2, \quad x_k = (1 - S_k) \ [k = 3, 15]$$

where the summations in $S_3$ and $S_{15}$ extend over the states included in the earlier [1,2] 3-state and 15-state studies, respectively; *the c's refer to the present calculation*. A glance at Table 1 shows that in $S_3$, the summation runs over $k = 1, 2$ and 15 (correspond to the same basis states as in ref. [1]), whereas in $S_{15}$ it runs over $k = 1–8$ and 15–21 (correspond to the same basis states as in ref. [2]). The fractions $x_3$ and $x_{15}$ are taken as approximate measures of the respective fractional improve-ments in the present calculations over the 3-state and over the 15-state calculations *as far as the "size effect" is concerned*. As regards the electron correlation effect, we should be careful to interpret the results and we should remember that it is not only the large basis size but also the fulfillment of conditions (4.2) and (4.3) above is a prime necessity to bring out the dynamic correlation effects in the calculated results of the collision process. If the said conditions are not satisfied, the difference between the small-basis and the large-basis results simply display a "size effect" and not dynamic electron correlation. This is a subtle point which, if lost or ignored, can lead to the confusion and doubt. Comparing the present calculations with our earlier work [1,2] we find that at low energy the 'size effect' is very little (not serious) for the single charge transfer and the series can truncated early, while the 'size effect' is serious and significant for double electron capture, and the early truncation of the series is inappropriate and cannot be done.

After this preamble, we do need to examine whether the detailed dynamical results obtained in this study reveal any signature of dynamic electron correlation. Table 3 gives the numerical values of the differences of probabilities ($P_3$ - $P_{27}$) and ($P_{15}$ - $P_{27}$), where $P_k$s, k = 3, 15 and 27, are the resonant double electron capture probabilities obtained from the k-state calculations of the earlier works [1,2] and from the present work, respectively, with the corresponding values of $x_3$ and $x_{15}$ ($x_{27}$ = 0 by definition) at a few values of *l* spread over the rightmost peaks in the *P-l* curves in Figs.3. These peaks have chosen because they contribute the most to the total charge transfer cross-sections. Also, we take the quantities ($P_3$ - $P_{27}$) and ($P_{15}$ - $P_{27}$) — or, to be more exact, the nature of their variation with the impact parameter — to be a measure of the *total effect* of the configuration interaction in the present work, *including both the "size effect" and the dynamic correlation effect* (*if any*), as compared to the earlier works.

From what has been said above, the effect of dynamic electron correlation at different energies should be capable of being singled out from a plot of the differences $x_k$ - ($P_k$ - $P_{27}$) = $D_k$ (say) where k = 3,15, versus the impact parameter, in conjunction with the natures of ($P_k$ - $P_{27}$) (k = 3, 15) in Table 3. Figures 6A–E show these quantities at five energies 2, 5, 7.5, 15 and 25 keV/amu, the same as in Figs. 3. At the two lowest energies (2 and 5 keV/amu), there is very little qualitative difference between the two curves, whereas at higher energies a remarkable difference in the *b*-variation of the two curves sets in. It is seen to note the following features:

(i) The extrema of the curves coincide more or less with the peaks of the *P - l* curves of Figs. 3.

(ii) Towards high energies the extrema positions seem to satisfy the condition (4.3), whereas at the lower energies the extrema lie farther than allowed by the said condition. In this connection, it is notable that with the 27-state wavefunction including π3p and σ3p orbitals (both *g* and *u*), the effective 'size' of the collision system is larger than that in the 15-state calculation.

Table 3: Separating out different effects in configuration interaction. See text for details.

| l | $x_3$ | $P_3 - P_{27}$ | $x_{15}$ | $P_{15} - P_{27}$ |
|---|---|---|---|---|
| | | [2 keV/amu] | | |
| 1400 | 0.048 | 0 | 0.005 | 0.071 |
| 1500 | 0.058 | 0.024 | 0.002 | 0.078 |
| 1600 | 0.061 | 0.050 | 0.001 | 0.068 |
| 1700 | 0.061 | 0.061 | 0.001 | 0.049 |
| 1800 | 0.070 | 0.053 | 0.001 | 0.020 |
| 1900 | 0.087 | 0.038 | 0.001 | 0 |
| | | [5 keV/amu] | | |
| 1800 | 0.114 | 0.053 | 0.021 | 0 |
| 1900 | 0.106 | 0.071 | 0.012 | 0 |
| 2000 | 0.097 | 0.080 | 0.008 | -0.004 |
| 2100 | 0.084 | 0.072 | 0.005 | -0.014 |
| 2200 | 0.065 | 0.062 | 0.005 | -0.026 |
| 2300 | 0.047 | 0.043 | 0.004 | -0.046 |
| 2400 | 0.039 | 0.030 | 0.004 | -0.055 |
| 2500 | 0.040 | 0.022 | 0.004 | -0.058 |
| 2600 | 0.047 | 0.016 | 0.004 | -0.058 |
| 2700 | 0.057 | 0.018 | 0.004 | -0.051 |
| 2800 | 0.065 | 0.018 | 0.004 | -0.043 |
| 2900 | 0.067 | 0.017 | 0.004 | -0.035 |
| 3000 | 0.062 | 0.017 | 0.004 | -0.028 |
| | | [7.5 keV/amu] | | |
| 2000 | 0.113 | 0.078 | 0.036 | -0.070 |
| 2200 | 0.066 | 0.068 | 0.018 | -0.104 |
| 2400 | 0.056 | 0.061 | 0.008 | -0.112 |
| 2600 | 0.059 | 0.063 | 0.005 | -0.084 |
| 2800 | 0.055 | 0.058 | 0.003 | -0.060 |
| 3000 | 0.044 | 0.042 | 0.003 | -0.052 |
| 3200 | 0.040 | 0.022 | 0.004 | -0.054 |
| 3400 | 0.049 | 0.008 | 0.006 | -0.054 |
| 3600 | 0.068 | 0.004 | 0.008 | -0.061 |
| | | [15 keV/amu] | | |
| 2000 | 0.295 | 0.139 | 0.116 | -0.016 |
| 2400 | 0.217 | 0.183 | 0.091 | -0.195 |
| 2800 | 0.164 | 0.158 | 0.049 | -0.309 |
| 3200 | 0.123 | 0.109 | 0.036 | -0.290 |
| 3600 | 0.098 | 0.071 | 0.025 | -0.272 |
| 4000 | 0.092 | 0.039 | 0.019 | -0.196 |
| 4400 | 0.087 | 0.016 | 0.016 | -0.145 |
| 4800 | 0.074 | 0.005 | 0.013 | -0.099 |
| 5400 | 0.058 | 0.005 | -0.005 | -0.055 |
| | | [25 keV/amu] | | |
| 2000 | 0.539 | 0.247 | 0.230 | -0.039 |
| 2400 | 0.445 | 0.286 | 0.179 | -0.124 |
| 2800 | 0.386 | 0.294 | 0.132 | -0.226 |
| 3200 | 0.330 | 0.268 | 0.104 | -0.305 |
| 3600 | 0.282 | 0.247 | 0.082 | -0.322 |
| 4000 | 0.237 | 0.202 | 0.059 | -0.311 |
| 4400 | 0.197 | 0.155 | 0.043 | -0.275 |
| 4800 | 0.165 | 0.109 | 0.029 | -0.228 |
| 5200 | 0.150 | 0.074 | 0.021 | -0.187 |
| 5600 | 0.151 | 0.031 | 0.019 | -0.148 |
| 6000 | 0.100 | 0.011 | 0.020 | -0.112 |

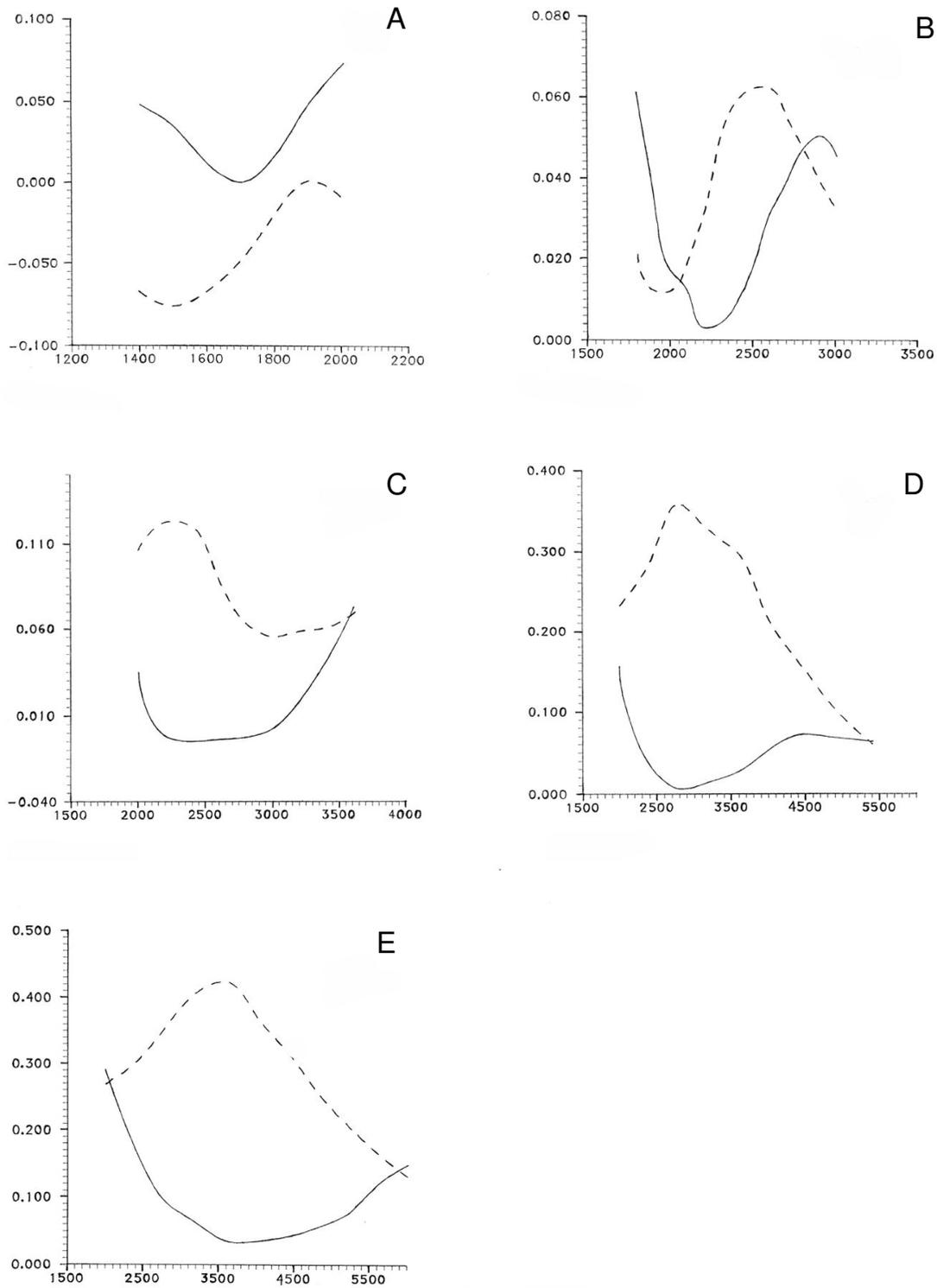

**Figure 6**
**(A-E). Full lines show $D_3$ and dashed lines show $D_{15}$ (see text) against *l* at the same energies (2, 5, 7.5, 15 and 25 keV/amu), respectively, as the corresponding Figs. 3**.

(iii) The impact velocities at the higher energies satisfy the condition(4.2) better than those at the lower energies do.

From these features one can find that effects of increase in the basis size at the low energies are contained almost completely in the "size effect"; a glance at Table 3 shows that this effect is almost saturated for the 15-state calculations. However, at the higher energies although the convergence (towards saturation) appeared in the 15-state results, i.e., columns $x_{15}$, in Table 3, the curves in Figs.6 indicate a distinct "overshooting" (peaked at around $l$ = 2300, 2600 and 3800 at 7.5, 15 and 25 keV/amu energies, respectively) due to the size effect in the 15-state calculations. The minima (negative values) of the total effect of the configuration interaction including the 'size effect' and 'dynamic electron correlation effect', $(P_{15} - P_{27})$ appeared at around $l$ = 2300, 3200 and 3800 at 7.5, 15 and 25 keV/amu energies, respectively, in Table 3, and $x_{27}$ = 0 by definition, this explains the 'size effect' in the 15-state calculations is "compensated for" in the 27-state calculations, by the dynamic electron correlation. Considering the *P-b* curves (i.e., *P* - *l* curves) in Figs.3, the 3-state results simply mimic the behaviour of single-electron capture of two electrons, as if the two electrons are moving independently; but other results show distinct dynamic correlation effects between the two electrons. (See the Appendix for a more detailed and quantitative exposition). Besides, this becomes more obvious when one compares the Figs.3 with the corresponding Figs.6. If we study the Figs. 3A and 3C, we find that the velocity in the latter is nearly twice that in the former, both being less than the orbital electron velocity (which is of order unity in a.u.). Note that although $l$ ~1650 in Fig. 3A and $l$ ~ 3000 in Fig. 3C correspond nearly to the same impact parameter, a comparison of Figs. 6A and 6C shows that *it is only in the figures* (C) *and not in the figures* (A) that the difference between the 3-state and the 27-state results can be attributed to dynamic correlation. This subtle point can be verified by comparing the other figures – Fig. 3B and 3E, at $l$ ~ 2000 and $l$ ~ 4000 respectively. Also, in figs. 3D and 3E we find a large difference between the 3-state peak and the 15-state peak around $l$ ~ 4000 but recognize it more likely to be a basis-set size effect than a signature of dynamic electron correlation, because condition (4.3) is not well satisfied here for these two basis sets. On the other hand, with 27-state wavefunction including diffuse orbitals, the effective "size" of the collision complex is larger and correspondingly closer to the impact parameter, this satisfying condition (4.3); as such, the difference between the 3-state and the 27-state results in this two figures does indicate signature of the dynamic electron correlation.

Moreover, we still have to answer the question: Does the double charge transfer experiment show any evidence of electron correlation, as indicated in this theoretical study? The answer that is provided by the total double charge transfer cross-section data is fairly definitive in this regard. It has been shown in the last paragraph how the double charge transfer probability values (shown in Figs.3) towards large $l$ – *the region which contributes the most to the total cross-section* – are affected by dynamic electron correlation when large basis sets are used. The fact that towards the higher energies the 3-state, 15-state and 27-state calculations yield progressively better agree-ments with experiment as shown in Fig.4, coupled with the above feature brought out in the Figs.3, provides numerical evidence that dynamic electron correlation seems to play an important role in the resonant two-electron transfer channel. As already mentioned, at the lowest energies (~1–4 keV/amu) the differences between the three sets of results reflect merely the basis-set size effect and cannot be construed to testify to dynamic electron correlation. Noteworthy that although a slight discrepancy with experiment persists in the energy region above 10 keV/amu, its magnitude is much less than that for the 15-state results. An odd feature, namely that the discrepancy has the *opposite sign*, seems to indicate that as one goes on increasing the size of the diabatic basis in the close-coupling approach, the proper (asymptotic) amount of CI that is required to fully incorporate the electron correlation is approached in an *oscillatory fashion*. (This feature also appears in the single capture results as shown in Fig.2, as mentioned above in Section 3.) An earlier study of electron impact ionization of He carried out by Bray and Fursa [15] also displays this oscillatory convergence of close-coupled calculations. Also, at these energies, Kimura's [10] results are bracketed by Afrosimov's data on the one side and by present 27-state results on the other. This seems convincing that, up to energies of $\approx$ 30 keV/amu, i.e. for velocities up to $\approx$ 1.1 a.u., unless we require benchmark accuracy in the results, electron translational factors would *not* need to be included in present kind of diabatic molecular basis where the configuration interaction has been properly included to investigate two-electron ion-atom collision system. This agrees with the observation of Zygelman *et al*.[16] that 'a molecular state expansion without electron translation factors is a valid low-energy approximation', mentioned in [2] as well.

It is expected that if the two electronic charges (-2$e$) are taken together as an equivalent single 'pseudo-electron' of charge -$q$ = -2$e$ (-$e$ is the real electronic charge and -$q$ is the equivalent doubly charged single 'pseudo-electron'), and the corresponding mass corrections are done for interaction Hamiltonians and rotational coupling matrix elements, the whole process of resonant double electron capture should be possible to be studied by carrying out a 'one-pseudo-electron' calculation. In this case the doubly charged single 'pseudo-electron' of charge -$q$ is exchanged from He atom to He$^{q+}$ ion [$He_A(1s^2) + He_B^{q+} \rightarrow He_A^{q+} + He_B(1s^2)$], i.e., correlated motion of two electrons, and one should get direct results of pure simultaneous two-electron transfer probabilities and hence the cross sections at different energies. The above illustration would provide another probable way to investigate the dynamic two-electron correlation effects in ion-atom charge transfer collisions (planning to apply this to the *same $He_A$ + $He_B^{2+}$* system at low energies in future).

## 5. Conclusion

Comparison of a 27-state close-coupled calculation of *He$^{2+}$-He* charge transfer collision with earlier 3-state and 15-state calculations suggests that (i) dynamic electron correlation effects become visible as the basis set is progressively increased, i.e., as the configuration interaction is taken into account more properly, provided the mentioned conditions in section-4 are satisfied, and (ii) present results provide the importance of these correlation effects in the resonant two-electron transfer at low (but not too low) energies. Also, for the diabatic molecular basis representation used in this work which go over to the correct separated-atom energy limits, the results indicate that electron translation factors (ETFs) may *not* need to be included in charge transfer collision studies in this energy range, unless we need benchmark accuracy. However, if ETFs are properly incorporated into the charge transfer collision studies at low energies, these diabatic molecular basis states can be used for benchmark calculations. In the energy range under consideration, the static ground state correlation as given in the Wannier picture seems to connect with and would go over into dynamic electron correlation in doubly charged ion-atom collision processes under suitable conditions. The present work may shed light on ion-atom interactions and dynamic two-electron correlation pictures of the charge transfer ion-atom collisions at low energies.

## Appendix

The interpretation here, that the 3-state results in Figs. 3A–E mimicking a resonant single-electron transfer behaviour indicate independent and not correlated motion of the electrons, is derived here in a more quantitative fashion. For the former case the probability amplitudes for transfer of the two individual electrons are uncorrelated, and one may write the joint two-electron transfer probability in the form

$$P(1,2) = |a_1 a_2|^2 = |a_1|^2 |a_2|^2 = P(1)\, P(2),$$

where $a_1$ and $a_2$ are the respective (normalized) transfer amplitudes for this resonant case and $P(1)$ and $P(2)$ are the individual transfer probabilities; by symmetry they are equal, and hence $P(1,2)$ would show the same features as they. On the other hand, for dynamically correlated motion the two-electron transfer amplitude is no longer given by the product of the two individual amplitudes $a_1$, $a_2$, but rather by a *correlated transfer amplitude* $a_{1,2}$, which in general *agrees neither in amplitude nor in phase with the said product*. The traces of the phase effects might be involved and might show up in the probabilities.

A comparison of the figures 3A–E shows that as we travel up energy and impact parameter, so that the conditions (4.2) and (4.3) are better satisfied (see the last two paragraphs in Sec. 4), the 27-state results for the joint two-electron resonant transfer probabilities show distinctly both these aspects of correlation effects (amplitude and phase mismatch) as compared to the 3-state results; the latter mimic the "$\sin^2$" behaviour of a resonant single-electron transfer and pass very much like a candidate for the uncorrelated $P(1,2)$ above.

### Acknowledgements
Computational facilities in the Department of Materials Science, Indian Association for the Cultivation of Science (IACS), Kolkata (Calcutta)- 32, used for the present work is acknowledged. I am grateful to Prof. T. K. Rai Dastidar, Dept. of Materials Science, IACS, Calcutta-32, India, for some helpful discussions. This work is dedicated to the memory of Prof. (Late) T. K. Rai Dastidar. The present work is also dedicated to the memory Prof.(Late) Mineo Kimura who has great contributions to ion-atom/molecule collision field. One of the three reviewers of this paper has dedicated his review report to the memory of Prof. M. Kimura.